\documentclass[11pt]{elsarticle}
\usepackage{geometry}
\usepackage{graphicx}

\usepackage{epsfig}
\usepackage{epstopdf}		

\usepackage{amsfonts,amsmath,amssymb,amscd}

\usepackage[figuresright]{rotating}

\usepackage{caption}
\usepackage{subcaption}

\usepackage[none]{hyphenat}

\usepackage{lineno}

\biboptions{round,sort&compress}

\newcommand{\pf}{{\bf Proof : }}
\newtheorem{definition}{Definition}
\newtheorem{theorem}{Theorem}[section]
\newtheorem{lemma}{Lemma}[section]

\newtheorem{example}{Example}[section]

\newcommand{\F}{\mathbb{F}}
\newcommand{\Z}{\mathbb{Z}}

\newcommand{\xn}{x^n - 1}

\begin{document}
	
\begin{frontmatter}

\title{DNA Cyclic Codes Over The Ring $ \F_2[u,v]/\langle u^2-1,v^3-v,uv-vu \rangle$ }

\author{Sukhamoy Pattanayak, Abhay Kumar Singh$^*$ and Pratyush Kumar}

\address{Department of Applied Mathematics, Indian School of Mines Dhanbad, India\\
	Email: sukhamoy88@gmail.com\\
	$^*$singh.ak.am@ismdhanbad.ac.in\\
	vikey397@gmail.com}

\begin{abstract}
In this paper, we mainly study the some structure of cyclic DNA codes of odd length over the ring $R=\F_2[u,v]/\langle u^2-1,v^3-v,uv-vu \rangle$ which play an important role in DNA computing. We established a direct link between the element of ring $R$ and 64 codons by introducing a Gray map from $R$ to $R_1=\F_2+u\F_2, u^2=1$ where $R_1$ is the ring of four elements. The reverse constrain and the reverse-complement constraint codes over $R$ and $R_1$ are studied in this paper. Binary image of the cyclic codes over $R$ also study. The paper concludes with some example on DNA codes obtained via gray map.
\end{abstract}


\end{frontmatter}
Mathematics Subject Classification  94B05. 94B15

\section{Introduction}
In algebraic coding theory Cyclic codes, an important class of linear codes over ring have recently generated a great deal of interest because of their rich algebraic structures and practical implementations. Cyclic codes over chain ring have been discussed by several Authors[[\cite {ab1}],[\cite{bo}]]. Zhu and Wang studied the algebraic structure of cyclic codes over $\F_2+v\F_2$, where $v^2=1$ in [\cite{zh1}]. In [\cite{yi3}], Linear Codes over $\F_2 +u\F_2 +v\F_2 + uv\F_2$ are studied. Yildiz and Karadeniz have considered the non chain ring $\F_2[u, v]/<u^2, v^2, uv - vu>$ and studied cyclic codes of odd length over that in [\cite{yi2}]. Cyclic codes over the ring $\Z_p[u, v]/<u^2, v^2, uv - vu>$ in [\cite{ke}] are studied by Kewat et al. In [\cite{sh}], Shi et al. discussed cyclic codes and weight enumerator linear codes over $\F_2+v\F_2+v^2\F_2$ where $v^3=v$. \\
Deoxyribonucleic acid (DNA) is a nucleic acid containing the genetic instructing used as the carrier of genetic information in all living organisms. DNA is formed by the strands and each strands is sequence consists of four nucleotides; two purines: adenine (A) and guanine (G), and two pyrimidines: thymine (T) and cytosine (C). The two strands of DNA are linked with a rule that are name as Watson-Crick complement (WCC). According to WCC rule; every (A) is linked with a (T), and every (C) with a (G), and vice versa. We write this is as $\overline{A}=T, \overline{T}=A, \overline{G}=C$ and $\overline{C}=G$. For example if $x=(GCATAG)$, then its complement is $\overline{x}=(CGTATC)$. \\
DNA computing links genetic data analysis with scientific computation in order to handle computationally difficult problems. Leonard Adleman [\cite{ad}] introduced an experiment involving the use of DNA molecules to solve a hard computational problem in a test tube. His study was based on the WCC property of DNA strands. Several paper have discussed different techniques to construct a set of DNA codewords that are unlikely to form undesirable bonds with each other by hybridization. Four different constraints on DNA codes are considered as follows:
\begin{enumerate}[{\rm (i)}]
	\item The \emph{Hamming constraint}: For any two codewords $x,y \in C, H(x,y)\geq d$ with $x\neq y$, for some minimum distance $d$.
	\item The \emph{reverse constraint}: For any two codewords $x,y \in C, H(x^r,y)\geq d$, where $x^r$ s the reverse of a codeword $x$.
	\item The \emph{reverse-complement constraint}: $ H(x^{rc},y)\geq d$ for all $x,y \in C$.
	\item The \emph{fixed GC-content constraint}: For any codeword $x \in C$ the same number of G and C elements.
\end{enumerate}
The constraints (i) to (iii) is to avoid undesirable hybridization between different strands. The fixed GC-content which ensures all codewords have similar thermodynamic characteristic.\\
Furthermore, cyclic DNA computing has generated great interest because of their more storage capacity than silicon based computing systems, and this motivates many authors to study it.
Since then, the construction of DNA cyclic codes have been discussed by several Authors in [[\cite{ab2}], [\cite{ba}], [\cite{be}], [\cite{ga}], [\cite{gu1}], [\cite{gu2}], [\cite{lia}], [\cite{si}], [\cite{yi1}]]. Gaborit and King in [\cite{ga}] discussed linear construction of DNA codes. In [\cite{ab2}], DNA codes over finite field with four elements were studied by Abualrub et al. Later, Siap et al. considered DNA codes over the finite ring $\F_2[u]/<u^2-1>$ with four element in [\cite{si}]. In [\cite{lia}], Liang and Wang discussed cyclic DNA codes over four element ring $\F_2+u\F_2$. Yildiz and Siap in [\cite{yi1}] show that the ring with 16 elements $\F_2[u]/<u^4-1>$ and DNA pairs matched for the first time and studied the algebraic structure of these DNA codes. Later in [\cite{ba}], codes over the ring $F_4+vF_4, v^2=v$ with 16 elements are considered by Bayram et al. and discussed some DNA application. Zhu and Chen studied cyclic DNA codes over the non chain ring $\F_2[u,v]/\langle u^2,v^2-v,uv-vu \rangle$ in [\cite{zh2}]. In [\cite{be}], Bennenni et al. considered the chain ring $\F_2[u]/<u^6>$ with 64 elements and discussed DNA cyclic code over this ring.  \\
Here we study a family of cyclic DNA codes of a ring $R=\F_2[u,v]/\langle u^2-1,v^3-v,uv-vu \rangle$ with 64 elements.
The sequence of paper is structured as follows: In the second section we discuss the structure of the ring $R=\F_2[u,v]/\langle u^2-1,v^3-v,uv-vu \rangle$ and $R_1=\F_2[u]\langle u^2-1,uv-vu \rangle$ and includes some basic background. We establish a 1-1 correspondence $\theta$ between 64 codons over the alphabet $\{A,T,G,C\}^3$ and the 64 elements of the ring $\F_2[u,v]/\langle u^2-1,v^3-v,uv-vu \rangle$ in section 3. Also we present a description and basic definition of cyclic DNA codes over the ring. We study cyclic codes satisfy the reverse constraint over $R_1$ and $R$ in section 3.1. In section 3.2, we also discuss cyclic codes satisfy the reverse-complement constraint over such rings. Moreover, we define the Lee weight related to such codes and give the binary image of the cyclic DNA code in section 4. In section 5, by applying the theory proved in the previous sections, we present some cyclic DNA codes over the ring $R$ and $R_1$ together with their images. Section 6 concludes the paper.

\section{Preliminaries}
Let $\F_2$ be the binary finite field. Throughout this paper $R$ be the commutative, characteristic 2 ring $\F_2+u\F_2+v\F_2+uv\F_2+v^2\F_2+uv^2\F_2=\{a_1+ua_2+va_3+uva_4+v^2a_5+uv^2a_6$, where $a_j \in \F_2, 1\leq j \leq 6 \}$ with $u^2=1, v^3=v$. $R$ can also be thought of as the quotient ring $ \F_2[u,v]/\langle u^2-1,v^3-v,uv-vu \rangle$.Let 
\begin{align}
R & = \F_2+u\F_2+v\F_2+uv\F_2+v^2\F_2+uv^2\F_2,~~~~\text{where}~~ u^2=1, v^3=v\nonumber \\ 
& =(\F_2+u\F_2)+v(\F_2+u\F_2)+v^2(\F_2+u\F_2), ~~~~\text{where}~~ u^2=1, v^3=v\nonumber \\ 
& = R_1+vR_1+v^2R_1, ~~~~\text{where}~~v^3=v.\nonumber 
\end{align}
Here $R_1$ be the finite chain ring $\F_2+u\F_2$ with $u^2=1$.\\
A linear code $C$ of length $n$ over $R$ is a $R$-submodule of $R^n$. An element of $C$ is called a codeword.
A code of length $n$ is cyclic if the code is invariant under the automorphism $\sigma$ which has
\begin{center}
	$\sigma(c_0, c_1, \cdots , c_{n-1}) = (c_{n-1}, c_0, \cdots , c_{n-2}).$
\end{center}
A code of length $n$ is 6-quasi cyclic if the code is invariant under the automorphism $\nu$ which has
\begin{center}
	$\nu(c_0, c_1, \cdots,c_{n-6},c_{n-5},c_{n-4},c_{n-3},c_{n-2},c_{n-1}) = (c_{n-6},c_{n-5},c_{n-4},c_{n-3},c_{n-2},c_{n-1}, c_0, \cdots , c_{n-7}).$
\end{center}
It is well known that a cyclic code of length $n$ over $R$ can be identified with an ideal in the quotient ring $R[x]/\langle x^n-1\rangle$ via the $R$-module isomorphism
as follows:
\begin{center}
	$R^n \longrightarrow R[x]/\langle x^n-1\rangle$\\
	$(c_0,c_1,\cdots,c_{n-1}) \mapsto c_0+c_1x+\cdots+c_{n-1}x^{n-1} (\text{mod}\langle x^n-1\rangle) $
\end{center}
Let $R_n=R[x]/\langle \xn \rangle,\text{and}~ R_{1,n}=R_1[x]/\langle \xn \rangle$. We assume that length $n$ is odd, and represent codewords by polynomials. Then cyclic codes are ideals in the ring $R_n$.\\
Since $R=R_1+vR_1+v^2R_1=R_1[v]/\langle v(1+v^2) \rangle$, for any element $x=a+bv+cv^2 \in R ~~\text{where}~a,b,c \in R_1$, according to the Chinese Remainder Theorem, $x=CRT^{-1}(a,a+b+c+bw),~~\text{where}~~w=v+v^2~~\text{and}~~w^2=0$. Similarly, 
\begin{center}
	$(a,A+Bw)\xrightarrow{CRT} a+Bw+(a+A+B)v^2 \in R,~~\text{where}~~A,B \in R_1$.
\end{center}		
Hence, a code $C$ over $R$ can be written as $C=CRT^{-1}(C_2,C_w),~~\text{where}~~C_2\in R_1, C_w\in R_1+wR_1$. $C_2$ is a cyclic code of odd length $n$ over $R_1$, then $C_2 = \langle g_2(x)+(1+u)a_2(x) \rangle$, where $a_2(x)\vert g_2(x) \vert (\xn)$. If $C_w$ is cyclic code of odd length $n$ over $R_w$, then $C_w =\langle g_1(x)+(1+u)a_1(x) + w(g_1^\prime(x)+(1+u)a_1^\prime(x))\rangle $, $a_1(x)\vert g_1(x) \vert (\xn)$, $a_1^\prime(x)\vert g_1^\prime(x) \vert (\xn)$.
The structure of cyclic code of arbitrary length $n$ over $R_1$ has been studied in [\cite{si}], which is
\begin{theorem}
	Let $C$ be a cyclic code in $R_{1,n}=R_1[x]/\langle \xn \rangle$. Then
	\begin{enumerate}[{\rm (1)}]
		\item	If $n$ is odd, then $R_{1,n}$ is a principal ideal ring and $C=\langle g(x), (1+u)a(x) \rangle$, where $g(x), a(x)$
		are binary polynomials with $a(x)\vert g(x) \vert (\xn)~\text{mod}~2$. 
		\item If $n$ is not odd, then
		\begin{enumerate}[{\rm (a)}]
			\item If $g(x)=a(x)$, then $C=\langle g(x)+(1+u)p(x) \rangle$, where $g(x), p(x)$
			are binary polynomials with $g(x) \vert (\xn)~\text{mod}~2$, $(g(x)+(1+u)p(x))\vert (\xn)~\text{in}~R_1$.
			\item $C=\langle g(x)+(1+u)p(x), (1+u)a(x) \rangle$, where $g(x), a(x)~\text{and}~p(x)$
			are binary polynomials with $a(x)\vert g(x) \vert (\xn)~\text{mod}~2$, $a(x)\vert p(x)((\xn)/g(x))$ and $~deg~g(x) > ~deg~a(x) > ~deg~p(x) $.
		\end{enumerate}
	\end{enumerate}	
\end{theorem}

\section{The reverse constraint and reverse-complement constraint codes over $R_1$ and $R$}
In this section, we study the reverse constraint and reverse-complement constraint codes over $R$ and $R_1$.  
DNA occurs in sequences, represented by sequences of nucleotides $S_{D_4}=\{A,T,G,C\}$. We define a DNA code of length $n$ to be a set of codewords $(x_0,x_1,\cdots, x_{n-1})$ where $x_i \in \{A,T,G,C\}$. These codewords must satisfy the four constraints mentioned above introduction. In this paper, the ring considered as 
$$R=\F_2[u,v]/\langle u^2-1,v^3-v,uv-vu\rangle=\{a_1+ua_2+va_3+uva_4+v^2a_5+uv^2a_6,  u^2=0,v^3=v \},$$
$~\text{where}~  a_j \in \F_2, 1\leq j \leq 6.$
Since the commutative ring $R$ is of the cardinality 64, then we define the map $\theta$ which gives a one-to-one correspondence between the elements of $R$ and the 64 codons over the alphabet $\{A,T,G,C\}^3$, which is given in Table 1. The elements $0,1,u,1 + u$
of $R_1=\F_2+u\F_2, u^2=1$ are in one-to-one correspondence with the nucleotide DNA bases $A,T,G,C$ such that
$0\rightarrow A, 1 \rightarrow G, u \rightarrow C ~\text{and}~ 1 + u \rightarrow T$. The codons satisfy the Watson-Crick complement which is given by $\overline{A}=T, \overline{T}=A, \overline{G}=C, \overline{C}=G$. \\
The Gray map $\Phi$ from $R$ to $R_1^3$ defined as follows
$$\Phi:R(=R_1+vR_1+v^2R_1, v^3=v)\longrightarrow R_1^3$$ such that~~~~~~~~~~~~~~~~ $\Phi(a+vb+v^2c)=(a,a+b,a+c)~\text{where}~~a,b,c \in R_1$.
\begin{definition}
	Let $C$ be a code over $R$ of arbitrary length $n$ and $c \in C$ be a codeword where $c=(c_0,c_1,\cdots,c_{n-1}), c_i \in R$, then we define 
	$$\Phi(c):C\longrightarrow S_{D_4}^{3n},$$ $(a_0+vb_0+v^2c_0,a_1+vb_1+v^2c_1,\cdots,a_{n-1}+vb_{n-1}+v^2c_{n-1})\longrightarrow (a_0,a_1,\cdots,a_{n-1},a_0+b_0,a_1+b_1,\cdots,a_{n-1}+b_{n-1},a_0+c_0,a_1+c_1,\cdots,a_{n-1}+c_{n-1})$.
\end{definition}
\newpage

\begin{center}
	{\bf Table 1.} Identifying Codons with the Elements of the Ring and Gray image $R$.\\~\\
	\begin{tabular}{c c c}
		\hline
		Elements 
		of $R$ & Gray image & DNA Codons \\
		\hline
		$0$ & $(0,0,0)$ & $AAA$ \\
		$v^2$ & $(0,0,1)$ & $AAG$  \\
		$uv^2$ & $(0,0,u)$ & $AAC$ \\
		$v^2+uv^2$ & $(0,0,1+u)$ & $AAT$   \\
		
		$v$ & $(0,1,0)$ & $AGA$  \\
		$v+v^2$ & $(0,1,1)$ & $AGG$   \\
		$v+uv^2$ & $(0,1,u)$ & $AGC$  \\
		$v+v^2+uv^2$ & $(0,1,1+u)$ & $AGT$   \\
		
		$uv$ & $(0,u,0)$ & $ACA$  \\
		$uv+v^2$ & $(0,u,1)$ & $ACG$  \\
		$uv+uv^2$ & $(0,u,u)$ & $ACC$  \\
		$uv+v^2+uv^2$ & $(0,u,1+u)$ & $ACT$    \\
		
		$v+uv$ & $(0,1+u,0)$ & $ATA$  \\
		$v+uv+v^2$ & $(0,1+u,1)$ & $ATG$  \\
		$v+uv+uv^2$ & $(0,u,u)$ & $ATC$ \\
		$v+uv+v^2+uv^2$ & $(0,u,1+u)$ & $ATT$ \\

			$u$ & $(u,u,u)$ & $CCC$ \\
			$u+v^2$ & $(u,u,1+u)$ & $CCT$  \\
			$u+uv^2$ & $(u,u,0)$ & $CCA$  \\
			$u+v^2+uv^2$ & $(u,u,1)$ & $CCG$   \\
			
		$u+v$ & $(u,1+u,u)$ & $CTC$ \\
		$u+v+v^2$ & $(u,1+u,1+u)$ & $CTT$   \\
		$u+v+uv^2$ & $(u,1+u,0)$ & $CTA$   \\
		$u+v+v^2+uv^2$ & $(u,1+u,1)$ & $CTG$   \\
			
			$u+uv$ & $(u,0,u)$ & $CAC$ \\
			$u+uv+v^2$ & $(u,0,1+u)$ & $CAT$   \\
			$u+uv+uv^2$ & $(u,0,0)$ & $CAA$   \\
			$u+uv+v^2+uv^2$ & $(u,0,1)$ & $CAG$  \\
			
			$u+v+uv$ & $(u,1,u)$ & $CGC$ \\
			$u+v+uv+v^2$ & $(u,1,1+u)$ & $CGT$   \\
			$u+v+uv+uv^2$ & $(u,1,0)$ & $CGA$   \\
			$u+v+uv+v^2+uv^2$ & $(u,1,1)$ & $CGG$   \\
		\hline
		
	\end{tabular}
\end{center}
\begin{center}
	{\bf Table 1.} Identifying Codons with the Elements of the Ring and Gray image $R$.\\~\\
	\begin{tabular}{c c c}
		\hline
		Elements of $R$ & Gray image & DNA Codons \\
		\hline
		$1$ & $(1,1,1)$ & $GGG$ \\
		$1+v^2$ & $(1,1,0)$ & $GGA$  \\
		$1+uv^2$ & $(1,1,1+u)$ & $GGT$  \\
		$1+v^2+uv^2$ & $(1,1,u)$ & $GGC$  \\
		
		$1+v$ & $(1,0,1)$ & $GAG$ \\
		$1+v+v^2$ & $(1,0,0)$ & $GAA$  \\
		$1+v+uv^2$ & $(1,0,1+u)$ & $GAT$ \\
		$1+v+v^2+uv^2$ & $(1,0,u)$ & $GAC$  \\
		
		$1+uv$ & $(1,1+u,1)$ & $GTG$ \\
		$1+uv+v^2$ & $(1,1+u,0)$ & $GTA$ \\
		$1+uv+uv^2$ & $(1,1+u,1+u)$ & $GTT$ \\
		$1+uv+v^2+uv^2$ & $(1,1+u,u)$ & $GTC$  \\
		
		$1+v+uv$ & $(1,u,1)$ & $GCG$ \\
		$1+v+uv+v^2$ & $(1,u,0)$ & $GCA$ \\
		$1+v+uv+uv^2$ & $(1,u,1+u)$ & $GCT$ \\
		$1+v+uv+v^2+uv^2$ & $(1,u,u)$ & $GCC$  \\

		$1+u$ & $(1+u,1+u,1+u)$ & $TTT$ \\
		$1+u+v^2$ & $(1+u,1+u,u)$ & $TTC$  \\
		$1+u+uv^2$ & $(1+u,1+u,1)$ & $TTG$  \\
		$1+u+v^2+uv^2$ & $(1+u,1+u,0)$ & $TTA$  \\
		
		$1+u+v$ & $(1+u,u,1+u)$ & $TCT$\\
		$1+u+v+v^2$ & $(1+u,u,u)$ & $TCC$  \\
		$1+u+v+uv^2$ & $(1+u,u,1)$ & $TCG$  \\
		$1+u+v+v^2+uv^2$ & $(1+u,u,0)$ & $TCA$  \\
		
	    $1+u+uv$ & $(1+u,1,1+u)$ & $TGT$\\
		$1+u+uv+v^2$ & $(1+u,1,u)$ & $TGC$  \\
		$1+u+uv+uv^2$ & $(1+u,1,1)$ & $TGG$  \\
		$1+u+uv+v^2+uv^2$ & $(1+u,1,0)$ & $TGA$  \\
		
		$1+u+v+uv$ & $(1+u,0,1+u)$ & $TAT$\\
		$1+u+v+uv+v^2$ & $(1+u,0,u)$ & $TAC$  \\
		$1+u+v+uv+uv^2$ & $(1+u,0,1)$ & $TAG$  \\
		$1+u+v+uv+v^2+uv^2$ & $(1+u,0,0)$ & $TAA$  \\
		\hline
		
	\end{tabular}
	\end{center}
Let $x=x_0x_1 \cdots x_{n-1} \in R^n$ be a vector. The reverse of $x$ is defined as $x^r = x_{n-1}x_{n-2} \cdots x_1x_0$, the complement of $x$ is $x^c =\overline{x_0}~\overline{x_1}\cdots \overline{x_{n-1}} $, and the reverse-complement, also called the Watson-Crick complement (WCC) is defined as $x^{rc} =\overline{x_{n-1}}~\overline{x_{n-2}} \cdots \overline{x_1}~\overline{x_0} $.
\begin{definition}
	A linear code $C$ of length $n$ over $R$ is said to be \bfseries{reversible} if $x^r \in C~~ \forall~ x \in C$, \bfseries{complement} if $x^c \in C~~ \forall~ x \in C$ and \bfseries{reversible-complement} if $x^{rc} \in C~~ \forall~ x \in C$.	
\end{definition}
\begin{definition}
	A cyclic code $C$ of length $n$ is called DNA code over $R$ if
	\begin{enumerate}[{\rm (1)}]
		\item $C$ is cyclic code, i.e. $C$ is an ideal of $R[x]/\langle x^n-1\rangle$;
		\item For any codeword $x \in C, x \neq x^{rc}$	and $x^{rc} \in C$.
	\end{enumerate}
\end{definition}
For each polynomial $f(x)=a_0+a_1x+\cdots+a_rx^r$ with $a_r\neq 0$, we define the reciprocal of $f(x)$ to be the polynomial
\begin{center}
	$f^*(x)=x^rf(1/x)=a_rx^r+a_{r-1}x+\cdots+a_0x^r$.
\end{center}
It is easy to see that $~deg~(f ^*(x)) \leq deg~(f (x))$ and if $a_0 \neq 0$, then $~deg~(f ^*(x)) = deg~(f (x))$. $f(x)$ is called a self-reciprocal polynomial if there is a constant $m$ such that $f^*(x) = mf(x)$.
\begin{lemma} \cite{gu2}
	Let $f(x),g(x)$ be any two polynomials in $R$ with $~deg~f(x) \geq ~deg~g(x)$. Then
	\begin{enumerate}[{\rm (1)}]
		\item $[f(x)g(x)]^*=f^*(x)g^*(x)$;
		\item $[f(x)+g(x)]^*=f^*(x)+x^{~deg~f-deg~g}g^*(x)$.
	\end{enumerate}
\end{lemma}

\subsection{\bfseries{The reverse constraint codes}}
In this section, we study the reverse constraint on cyclic codes over $R_1$ and $R$. First, we give lemma which use the following theorems.

\begin{lemma} \cite{mas}
	Let $C = \langle f (x)\rangle$ be a cyclic code over $\F_2$, then $C$ is reversible if and
	only if $f(x)$ is self-reciprocal.
\end{lemma}
First we study the reverse constraint on cyclic code of arbitrary length over $R_1$.
\begin{theorem} \cite{si}
	Let $C = \langle g(x), (1+u)a(x) \rangle = \langle g(x) + (1+u)a(x)\rangle$ be a cyclic code of odd length $n$ over $R_1$. Then $C$ is reversible if and only if $g(x)$ and $a(x)$ are self-reciprocal.
\end{theorem}
\begin{theorem}
	Let $C=\langle g(x) + (1+u)p(x)\rangle$ be a cyclic code of even length $n$ over $R_1$. Then $C$ is reversible if and only if
	\begin{enumerate}[{\rm (a)}]
		\item $g(x)$ is self-reciprocal;
		\item \begin{enumerate}[{\rm (i)}]
	    \item $x^ip^*(x)=p(x)$ and
	    \item $g(x) = x^i p^∗(x) + p(x)$, 
			where $i=~deg~g(x)-deg~p(x)$.
		\end{enumerate}
	\end{enumerate}
\end{theorem}
\pf Suppose $C=\langle g(x) + (1+u)p(x)\rangle$ is reversible over $R_1$, then $C$ $~mod~ 1+u$ is reversible over $\F_2$, then from Lemma 3.2, $g(x)$ is self-reciprocal. That implies 
\begin{align}
	(g(x) + (1+u)p(x))^* & = g^*(x)+ (1+u)x^ip^*(x)\nonumber \\ 
	& = g(x)+(1+u)x^ip^*(x)\\
	& = (g(x)+(1+u)p(x))a(x)
\end{align}
where $i=~deg~g(x)-deg~p(x)$. Comparing the degree of (1) and (2), we get $a(x)=c$, where $c\in R_1$. Then
\begin{equation}
	g(x)+(1+u)x^ip^*(x)=c.g(x)+c.(1+u)p(x).
\end{equation}
Hence $(1+u)g(x)=c.(1+u)g(x)$. That implies $c=1 ~\text{or}~u$. If $c=1$, then $x^ip^*(x)=p(x)$. For $c=u$, we get $g(x) = x^i p^∗(x) + p(x)$. 

On the other hand
\begin{align}
	(g(x) + (1+u)p(x))^* & = g^*(x)+(1+u)x^ip^*(x)\nonumber \\ 
	& = g(x)+(1+u)x^ip^*(x)\nonumber \\
	& = (g(x) + (1+u)p(x))c \in C, \nonumber
\end{align}
where $c=1 ~\text{or}~u$, $i=deg~g(x)-deg~p(x)$. Hence, $C$ is reversible.
\begin{theorem}
	Let $C=\langle g(x) + (1+u)p(x),(1+u)a(x) \rangle$
	with $a(x)\vert g(x)\vert \xn, a(x)\vert p(x)  (\frac{x^n-1}{g(x)})$ and $deg~a(x)>deg~p(x)$ be a cyclic code of even length $n$ over $R_1$. Then $C$ is reversible if and only if
	\begin{enumerate}[{\rm (a)}]
		\item $g(x)$ and $a(x)$ are self-reciprocal;
		\item $a(x)\vert (x^ip^*(x)+p(x))$ ,
			where $i=~deg~g(x)-deg~p(x)$.
	\end{enumerate}
\end{theorem}
\pf Suppose $C=\langle g(x) + (1+u)p(x),(1+u)a(x) \rangle $ is reversible. Then $\langle g(x)\rangle$ and  $\langle a(x)\rangle$ are reversible over $\F_2$, from Lemma 3.2 $\langle g(x)\rangle$ and  $\langle a(x)\rangle$ are self-reciprocal. Since $C$ is reversible this implies
\begin{align}
	(g(x) + (1+u)p(x))^* & = g^*(x)+ (1+u)x^ip^*(x)\nonumber \\ 
	& = g(x)+(1+u)x^ip^*(x)\\
	& = (g(x)+(1+u)p(x))b_1(x)+(1+u)a(x)b_2(x),
\end{align}
where $i=~deg~g(x)-deg~p(x)$. Comparing the degree of (4) and (5), we have $b_1(x)=c$, where $c\in R$. Then
\begin{equation}
	g(x)+(1+u)x^ip^*(x)=c.g(x)+c.(1+u)p(x)+(1+u)a(x)b_2(x).
\end{equation}
Multiplying $1+u$ both side of (6) write, $(1+u)g(x)=c.(1+u)g(x)$. That implies $c=1 ~\text{or}~u$. If $c = 1$, we have $(1+u)x^i p^∗(x)+(1+u)p(x) = (1+u)a(x)b_2(x)$. Which means $x^i p^∗(x)+p(x) \in (a(x))$. Therefore, $a(x)\vert (x^i p^∗(x)+p(x))$. For $c = u$, we
have $(1+u)g(x)+(1+u)x^i p^∗(x)+(1+u)p(x) = (1+u)a(x)b_2(x)$, Hence $g(x)+x^i p^∗(x)+p(x) \in (a(x))$.
Therefore, $a(x)\vert (x^i p^∗(x)+ p(x)+g(x))$. Since $a(x)\vert g(x)$, we have $a(x)\vert (x^i p^∗(x)+p(x))$. So in both cases, we have $a(x)\vert (x^i p^∗(x)+p(x))$.

Conversely, for $C$ to be reversible it is sufficient to show that both $(g(x) + (1+u)p(x))^*$ and $((1+u)a(x))^*$ are in $C$. Since $a(x)$ is self-reciprocal then $(1+u)a^*(x) = (1+u)a(x) \in C$. Also
\begin{align}
(g(x) + (1+u)p(x))^* & = g^*(x)+ (1+u)x^ip^*(x)\nonumber \\ 
& = g(x)+(1+u)x^ip^*(x)\nonumber \\
& = (g(x)+(1+u)p(x))+(1+u)p(x)+(1+u)p^*(x) \in C.\nonumber
\end{align}
Therefore, $C$ is reversible.\\
We will give one of the main result below.
\begin{lemma}
	Let $C$ be a cyclic code of odd length $n$ over $R$. Then $C$ is an ideal in $R_n$, which can be generated by $C = CRT^{-1}(C_2,C_w)$, where
	$C_2 = \langle g_2(x)+(1+u)a_2(x) \rangle$, $C_w =\langle g_1(x)+(1+u)a_1(x) + w(g_1^\prime(x)+(1+u)a_1^\prime(x))\rangle $, $a_2(x)\vert g_2(x) \vert (\xn)$, $a_1(x)\vert g_1(x) \vert (\xn)$, $a_1^\prime(x)\vert g_1^\prime(x) \vert (\xn)$.
\end{lemma}
\begin{theorem}
	Let $C=CRT^{-1}(C_2,C_w)$ be a cyclic code of odd length $n$ over $R$. Then $C$ is reversible if and only if $C_2$ and $C_w$ are reversible and degree of $g_2(x),g_1(x),g_1^\prime(x),a_2(x),a_1(x),a_1^\prime(x)$ are equal, where $C_2$ is cyclic code over $R_1$ and $C_w$ is cyclic code over $R_w$.
\end{theorem}
\pf If $C_2$ and $C_w$ are reversible, we have $C_2^r \in C$ and $C_w^r \in C$. Here $C_2 = \langle g_2(x)+(1+u)a_2(x) \rangle$ is reversible, that implies $g_2(x)$ and $a_2(x)$ are self reciprocal. That is $g_2^*(x)=g_2(x)$ and $a_2^*(x)=a_2(x)$. Again $C_w =\langle g_1(x)+(1+u)a_1(x) + w(g_1^\prime(x)+(1+u)a_1^\prime(x))\rangle $ is reversible, that implies $g_1(x), a_1(x), g_1^\prime(x)$ and $a_1^\prime(x)$ are self reciprocal. That is $g_1^*(x)=g_1(x), a_1^*(x)=a_1(x), {g_1^\prime}^*(x)=g_1^\prime(x)$ and ${a_1^\prime}^*(x)=a_1^\prime(x)$. For any $g(x) \in C, g(x)=g_2(x)+(1+u)a_2(x)+[g_1(x)+(1+u)a_1(x)]v+[g_2(x)+(1+u)a_2(x)+g_1^\prime(x)+(1+u)a_1^\prime(x)+g_1(x)+(1+u)a_1(x)]v^2$. Therefore
\begin{multline}
[g(x)]^* =[(1+v^2)g_2(x)+(1+u+v^2+uv^2)a_2(x)+(v+v^2)g_1(x)+(v+uv+v^2+uv^2)a_1(x)\\+v^2g_1^\prime(x)+(v^2+uv^2)a_1^\prime(x)]^*\\ = (1+v^2)g_2^*(x)+x^{deg~g_2-deg~a_2}(1+u+v^2+uv^2)a_2^*(x)+x^{deg~g_2-deg~g_1}(v+v^2)g_1^*(x) \\ +x^{deg~g_2-deg~a_1}(v+uv+v^2+uv^2)a_1^*(x)+x^{deg~g_2-deg~a_1^\prime}v^2{g_1^\prime}^*(x)+x^{deg~g_2-deg~a_1^\prime}(v+v^2){a_1^\prime}^*(x)\\=(1+v^2)g_2(x)+(1+u+v^2+uv^2)a_2(x)+(v+v^2)g_1(x)+(v+uv+v^2+uv^2)a_1(x)\\+v^2g_1^\prime(x)+(v^2+uv^2)a_1^\prime(x)=g(x). \nonumber
\end{multline}
As degree of $g_2(x),g_1(x),g_1^\prime(x),a_2(x),a_1(x),a_1^\prime(x)$ are equal.
Hence $g(x)$ is self reciprocal, that implies $C$ is reversible code.\\
On the other hand, if $C=CRT^{-1}(C_2,C_w)$ is reversible, then $g(x)$ is self reciprocal i.e.,$g^*(x)=g(x)$. That implies $g_2(x),a_2(x), g_1(x), a_1(x), g_1^\prime(x)$ and $a_1^\prime(x)$ are self reciprocal. Hence $C_2$ and $C_w$ are reversible.

\subsection{\bfseries{The reverse-complement constraint codes}}

In this section, cyclic codes over $R_1$ and $R$ satisfy the reverse-complement are examined.
First, we give some useful lemmas which can be easily check.
\begin{lemma}
	For any $a \in R$ we have $a+\overline{a}=1+u$.
\end{lemma}
\begin{lemma}
	For any $a,b\in R$, then $\overline{a+b}=\overline{a}+\overline{b}+(1+u)$.
		
\end{lemma}
\begin{lemma}
	For any $a \in \F_2$ we have $ (1+u)+\overline{(1+u)a}=(1+u)a$.
\end{lemma}

\begin{theorem}\cite{si}
	Let $C = \langle g(x), (1+u)a(x) \rangle$ be a cyclic code of length $n$ in $R_{1,n}$. If $f(x)^{rc} \in C$ for
	any $f(x)\in C$, then $(1+u)/((\xn) / (x-1))\in C$, and $g(x)$ and $a(x)$ are self-reciprocal polynomials.
\end{theorem}
\begin{theorem}\cite{si}
	Suppose $C = \langle g(x), (1+u)a(x) \rangle$ be a cyclic code of length $n$. If $(1+u)/((\xn) / (x-1))\in C$, and $g(x)$ and $a(x)$ are self-reciprocal polynomials, then $f(x)^{rc} \in C$ for
	any $f(x)\in C$.
\end{theorem}
We will give one of our main conclusions below.
\begin{theorem}
	Let $C=CRT^{-1}(C_2,C_w)$ be a cyclic code of odd length $n$ over $R$. Then $C$ is reversible-complement if and only if $C$ is reversible and $(\overline{0},\overline{0},\cdots,\overline{0})\in C$, where $C_2$ is cyclic code over $R_1$ and $C_w$ is cyclic code over $R_w$.
\end{theorem}
\pf Suppose $C=CRT^{-1}(C_2,C_w)$, where $C_2$ is cyclic code over $R_1$ and $C_w$ is cyclic code over $R_w$. For any $c = (c_0, c_1, \cdots, c_{n-1})\in C$, $c^{rc} = (\overline{c_{n-1}}, \overline{c_{n-2}}, \cdots, \overline{c_0})\in C$ as $C$ is reversible-complement. Since the zero codeword is in $C$ then its WCC is also in $C$, i.e.,
\begin{equation}
  (\overline{0},\overline{0},\cdots,\overline{0})\in C
\end{equation}
Whence, 
\begin{equation}
 c^r=(c_{n-1}, c_{n-2},\cdots,c_0)=(\overline{c_{n-1}}, \overline{c_{n-2}}, \cdots, \overline{c_0})+(\overline{0},\overline{0},\cdots,\overline{0})\in C.
\end{equation}
 ~~~~~~~~~~~~~On the other hand, if $C$ is reversible, then for any $c = (c_0, c_1, \cdots, c_{n-1})\in C$, $c^r=(c_{n-1}, c_{n-1},\cdots,c_0)\in C$. Since $(\overline{0},\overline{0},\cdots,\overline{0})\in C$, we get 
 \begin{equation}
  c^{rc} = (\overline{c_{n-1}}, \overline{c_{n-2}}, \cdots, \overline{c_0})=(c_{n-1}, c_{n-2},\cdots,c_0)+(\overline{0},\overline{0},\cdots,\overline{0})\in C.
 \end{equation}
Hence, $C$ is reversible-complement.

\section{Binary images of DNA codes over $R$}
In this Section we will define a Gray map which allows us to translate the properties of the suitable DNA codes for DNA computing to the binary cases. Now we define the Gray map on $R$. Any element $c \in R$ can be expressed as $c=a_1+ua_2+va_3+uva_4+v^2a_5+uv^2a_6$, where $a_j \in \F_2, 1\leq j \leq 6$. The Gray map $\phi$ defined as follows 
\begin{center}
	$\phi:R \longrightarrow \F_2^6$\\
	such that ~~~~ $\phi (a_1+ua_2+va_3+uva_4+v^2a_5+uv^2a_6)=(a_1, a_2, a_1+a_3,a_2+a_4,a_1+a_5,a_2+a_6)$,~~~~~~~~$a_j \in \F_2, 1\leq j \leq 6$.
\end{center}

The Lee weight was defined over the ring $R$ as follows 
\begin{center}
	$w_L(a_1+ua_2+va_3+uva_4+v^2a_5+uv^2a_6)=\displaystyle\sum _{i=1}^{6}a_i$.
\end{center}
 The Lee distance $d_L$, given by $d_L(c_1,c_2)=w_L(c_1-c_2)$. The Hamming distance $d(c_1, c_2)$ between two codewords $c_1$ and $c_2$ is the Hamming weight of the codeword $c_1-c_2$. It is easy to verify that the image of a linear code over $R$ by $\phi$ is a binary linear code. In Table 2 we give the
binary image of the codons. In [\cite{mi}] the binary image of DNA code resolved the problem of the construction of DNA codes with some properties. 
\begin{center}
	{\bf Table 2.} Binary Image of the Codons\\~\\
	\begin{tabular}{ l  c  c  c  c  c  c  c  c }
		\hline
		
		$AAA$ & $000000$ & $CCC$ & $010101$ & $GGG$ & $101010$ & $TTT$ & $111111$ \\
		$AAG$ & $000010$ & $CCT$ & $010111$ & $GGA$ & $101000$ & $TTC$ & $111101$ \\
		$AAC$ & $000001$ & $CCA$ & $010100$ & $GGT$ & $101011$ & $TTG$ & $111110$ \\
		$AAT$ & $000011$ & $CCG$ & $010110$ & $GGC$ & $101001$ & $TTA$ & $111100$ \\
		$AGA$ & $001000$ & $CTC$ & $011101$ & $GAG$ & $100010$ & $TCT$ & $110111$ \\
		$AGG$ & $001010$ & $CTT$ & $011111$ & $GAA$ & $100000$ & $TCC$ & $110101$ \\
		$AGC$ & $001001$ & $CTA$ & $011100$ & $GAT$ & $100011$ & $TCG$ & $110110$ \\
		$AGT$ & $001011$ & $CTG$ & $011110$ & $GAC$ & $100001$ & $TCA$ & $110100$ \\
		
		$ACA$ & $000100$ & $CAC$ & $010001$ & $GTG$ & $101110$ & $TGT$ & $111011$ \\
		$ACG$ & $000110$ & $CAT$ & $010011$ & $GTA$ & $101100$ & $TGC$ & $111001$ \\
		$ACC$ & $000101$ & $CAA$ & $010000$ & $GTT$ & $101111$ & $TGG$ & $111010$ \\
		$ACT$ & $000111$ & $CAG$ & $010010$ & $GTC$ & $101101$ & $TGA$ & $111000$ \\
		$ATA$ & $001100$ & $CGC$ & $011001$ & $GCG$ & $100110$ & $TAT$ & $110011$ \\
		$ATG$ & $001110$ & $CGT$ & $011011$ & $GCA$ & $100100$ & $TAC$ & $110001$ \\
		$ATC$ & $001101$ & $CGC$ & $011000$ & $GCT$ & $100111$ & $TAG$ & $110010$ \\
		$ATT$ & $001111$ & $CGG$ & $011010$ & $GCC$ & $100101$ & $TAA$ & $110000$ \\
		\hline
		
	\end{tabular}
\end{center}

The following property of the binary image of the DNA codes comes from the definition. 
\begin{lemma}
	The Gray map $\phi$ is a distance-preserving map from ($R^n$, Lee distance) to ($\F_2^{6n}$ , Hamming distance) and this map also $\F_2$ linear.
\end{lemma}
\pf
From the definitions, it is clear that $\phi(c_1-c_2)=\phi(c_1)-\phi(c_2)$ for $c_1,c_2 \in R^n$. Thus, $d_L(c_1,c_2)=w_L(c_1-c_2)=w_H(\phi(c_1-c_2))=w_H(\phi(c_1)-\Phi(c_2))=d_H(\phi(c_1),\phi(c_2))$. Let $c_1,c_2 \in R^n, k_1,k_2 \in \F_2$, then from the definition of the Gray map, we have\\ $\phi(k_1c_1+k_2c_2)=k_1\phi(c_1)+k_2\phi(c_2)$, that implies $\phi$ is $\F_2$ linear.

\begin{theorem}
	Let $\sigma$ be the cyclic shift of $R^n$ and $\nu$ denote the 6-QC shift of $\F_2^{6n}$. Let $\phi$ be the Gray map from $R^n$ to $\F_2^{6n}$. Then prove that $\phi \sigma=\nu \phi$
\end{theorem}
\pf
Let $c=(c_0,c_1,\cdots,c_{n-1})\in R^n$, where $c_i=a_{1i}+ua_{2i}+va_{3i}+uva_{4i}+v^2a_{5i}+uv^2a_{6i}$, with $a_{1i},a_{2i},a_{3i},a_{4i},a_{5i},a_{6i} \in \F_2, 0\leq i\leq n-1$. From the definition of the Gray map, we get \\
$\phi(c) = (a_{10},a_{20},a_{10}+a_{30},a_{20}+a_{40},a_{10}+a_{50},a_{20}+a_{60},a_{11},a_{21},a_{11}+a_{31},a_{21}+a_{41},a_{11}+a_{51},a_{21}+a_{61},\cdots,a_{1(n-1)},a_{2(n-1)},a_{1(n-1)}+a_{3(n-1)},a_{2(n-1)}+a_{4(n-1)},a_{1(n-1)}+a_{5(n-1)},a_{2(n-1)}+a_{6(n-1)}). $\\
Hence \\
$\nu (\phi(c))=(a_{1(n-1)},a_{2(n-1)},a_{1(n-1)}+a_{3(n-1)},a_{2(n-1)}+a_{4(n-1)},a_{1(n-1)}+a_{5(n-1)},a_{2(n-1)}+a_{6(n-1)},\\ a_{10},a_{20},a_{10}+a_{30},a_{20}+a_{40},a_{10}+a_{50},a_{20}+a_{60},\cdots,a_{1(n-2)},a_{2(n-2)},a_{1(n-2)}+a_{3(n-2)},a_{2(n-2)}+a_{4(n-2)},a_{1(n-2)}+a_{5(n-2)},a_{2(n-2)}+a_{6(n-2)})$. \\
On the other hand,\\ 
$\sigma(c)=(c_{n-1},c_0,c_1,\cdots,c_{n-2})$\\
we deduce that \\
$\phi(\sigma(c))=(a_{1(n-1)},a_{2(n-1)},a_{1(n-1)}+a_{3(n-1)},a_{2(n-1)}+a_{4(n-1)},a_{1(n-1)}+a_{5(n-1)},a_{2(n-1)}+a_{6(n-1)},\\ a_{10},a_{20},a_{10}+a_{30},a_{20}+a_{40},a_{10}+a_{50},a_{20}+a_{60},\cdots,a_{1(n-2)},a_{2(n-2)},a_{1(n-2)}+a_{3(n-2)},a_{2(n-2)}+a_{4(n-2)},a_{1(n-2)}+a_{5(n-2)},a_{2(n-2)}+a_{6(n-2)})$. \\
Therefore ~~~$\phi \sigma=\nu \phi$.

\begin{theorem}
	If $C$ is a cyclic DNA code of length $n$ over $R$, then $\phi(C)$ is a binary quasi-cyclic DNA code of the length $6n$ and of index 6.
\end{theorem}
\pf
Let $C$ be a cyclic DNA code of length $n$ over $R$. Hence $\phi(C)$ is a set of length $6n$ over the alphabet $\F_2$ which is a quasi-cyclic code of index 6.

\section{Example}
In this section, we give some examples of cyclic codes of different lengths over the ring $R$ and $R_1$ to illustrate the above results. 
\begin{example}
	Cyclic codes of length $3$ over $R = \F_2+u\F_2+v\F_2+uv\F_2+v^2\F_2+uv^2\F_2 , u^2 = 1, v^3=v$: We have
	$$ x^3-1 = (x+1)(x^2 + x +1)=f_1f_2 ~\text{over}~ \F_2.$$
	Suppose $g_2(x)=a_2(x)=g_1(x)=a_1(x)=g_1^\prime(x)=a_1^\prime(x)=(x+1)$. Let $C_2 = \langle g_2(x)+(1+u)a_2(x)$, $C_w =\langle g_1(x)+(1+u)a_1(x) + w(g_1^\prime(x)+(1+u)a_1^\prime(x))\rangle $. It is easy to check that 
	$g_2(x),a_2(x),g_1(x),a_1(x),g_1^\prime(x) ~\text{and}~a_1^\prime(x)$ are self-reciprocal. Since $C = CRT^{-1}(C_2,C_w)$ be a cyclic code, then we can obtain a code with generator 
	\begin{multline}
	g(x) =(x+1)+(1+u)(x+1)+v(x+1)+v(1+u)(x+1)+[(x+1)+(1+u)(x+1)+(x+1)\\+(1+u)(x+1)+(x+1)+(1+u)(x+1)]v^2 \nonumber \\=
	(u+uv+uv^2)+(u+uv+uv^2)x. \nonumber
	\end{multline}
	 Therefore, $g^*(x)=(u+uv+uv^2)x+(u+uv+uv^2)=g(x)$.	 By the theorem $3.4$, $C$ is reversible code of length 3 over $R$.
\end{example}
Let $C_2 = \langle g_2(x)+(1+u)a_2(x)$ be a cyclic code of length 3 over $R_1=\F_2+u\F_2, u^2=1$ and $C_w =\langle g_1(x)+(1+u)a_1(x) + w(g_1^\prime(x)+(1+u)a_1^\prime(x))\rangle $ be a cyclic code of length 3 over $R_w=R_1+wR_1, w^2=(v+v^2)^2=0$, where $g_2(x)=a_2(x)=g_1(x)=a_1(x)=g_1^\prime(x)=a_1^\prime(x)=f_2$. The image of $C = CRT^{-1}(C_2,C_w)$ under the gray map $\Phi$ is a DNA code of length $9$. This code has 64 codewords which are listed in the Table 3.
\begin{center}
	{\bf Table 3.} All 64 codewords of $C$\\~\\
	\begin{tabular}{ c  c  c  c }
		\hline
		
		$AAAAAAAAA$ &  $TTTTTTTTT$ &  $CCCCCCCCC$ &  $GGGGGGGGG$  \\
		$AAAAAAGGG$ &  $TTTTTTCCC$ &  $CCCCCCTTT$ &  $GGGGGGAAA$  \\
		$AAAAAACCC$ &  $TTTTTTGGG$ &  $CCCCCCAAA$ &  $GGGGGGTTT$  \\
		$AAAAAATTT$ &  $TTTTTTAAA$ &  $CCCCCCGGG$ &  $GGGGGGCCC$  \\
		
		$AAAGGGAAA$ &  $TTTCCCTTT$ &  $CCCTTTCCC$ &  $GGGAAAGGG$  \\
		$AAAGGGGGG$ &  $TTTCCCCCC$ &  $CCCTTTTTT$ &  $GGGAAAAAA$  \\
		$AAAGGGCCC$ &  $TTTCCCGGG$ &  $CCCTTTAAA$ &  $GGGAAATTT$  \\
		$AAAGGGTTT$ &  $TTTCCCAAA$ &  $CCCTTTGGG$ &  $GGGAAACCC$  \\
		
		$AAACCCAAA$ &  $TTTGGGTTT$ &  $CCCAAACCC$ &  $GGGTTTGGG$  \\
		$AAACCCGGG$ &  $TTTGGGCCC$ &  $CCCAAATTT$ &  $GGGTTTAAA$  \\
		$AAACCCCCC$ &  $TTTGGGGGG$ &  $CCCAAAAAA$ &  $GGGTTTTTT$  \\
		$AAACCCTTT$ &  $TTTGGGAAA$ &  $CCCAAAGGG$ &  $GGGTTTCCC$  \\
		
		$AAATTTAAA$ &  $TTTAAATTT$ &  $CCCGGGCCC$ &  $GGGCCCGGG$  \\
		$AAATTTGGG$ &  $TTTAAACCC$ &  $CCCGGGTTT$ &  $GGGCCCAAA$  \\
		$AAATTTCCC$ &  $TTTAAAGGG$ &  $CCCGGGAAA$ &  $GGGCCCTTT$  \\
		$AAATTTTTT$ &  $TTTAAAAAA$ &  $CCCGGGGGG$ &  $GGGCCCCCC$  \\
		\hline
		
	\end{tabular}
\end{center}

\begin{example}
	Cyclic codes of length $5$ over $R = \F_2+u\F_2+v\F_2+uv\F_2+v^2\F_2+uv^2\F_2 , u^2 = 1, v^3=v$: We have
	$$ x^5-1 = (x+1)(x^4 + x^3 + x^2 + x +1)=f_1f_2 ~\text{over}~ \F_2.$$
	Suppose $g_2(x)=a_2(x)=g_1(x)=a_1(x)=g_1^\prime(x)=a_1^\prime(x)=f_2$. Let $C_2 = \langle g_2(x)+(1+u)a_2(x)$, $C_w =\langle g_1(x)+(1+u)a_1(x) + w(g_1^\prime(x)+(1+u)a_1^\prime(x))\rangle $. It is easy to check that 
	$g_2(x),a_2(x),g_1(x),a_1(x),g_1^\prime(x) ~\text{and}~a_1^\prime(x)$ are self-reciprocal. Since $C = CRT^{-1}(C_2,C_w)$ be a cyclic code, then we can obtain a code with generator 
	\begin{multline}
	g(x) =(x^4 + x^3 + x^2 + x +1)+(1+u)(x^4 + x^3 + x^2 + x +1)+v(x^4 + x^3 + x^2 + x +1)+\\~~~~~v(1+u)(x^4 + x^3 + x^2 + x +1)+[(x^4 + x^3 + x^2 + x +1)+(1+u)(x^4 + x^3 + x^2 + x +1)+(x^4 + x^3 + x^2 + x +1)\\+(1+u)(x^4 + x^3 + x^2 + x +1)+(x^4 + x^3 + x^2 + x +1)+(1+u)(x^4 + x^3 + x^2 + x +1)]v^2 \nonumber \\=
	(u+uv+uv^2)+(u+uv+uv^2)x+(u+uv+uv^2)x^2+(u+uv+uv^2)x^3+(u+uv+uv^2)x^4. \nonumber
	\end{multline}
	Therefore, $g^*(x)=(u+uv+uv^2)x^4+(u+uv+uv^2)x^3+(u+uv+uv^2)x^2+(u+uv+uv^2)x+(u+uv+uv^2)x=g(x)$.	 By the theorem $3.4$, $C$ is reversible code of length 5 over $R$.
\end{example}

\begin{example}
	\begin{enumerate}[{\rm (i)}]
		\item Cyclic codes of length $8$ over $R_1 = \F_2 + u \F_2 , u^2 = 1$: We have
		$$ x^8-1 = (x+1)^8=f^8 ~\text{over}~ \F_2.$$ 
		Let $C=\langle g(x) + (1+u)p(x)\rangle, ~\text{where}~g(x)=f^4, p(x)=x^3+x $ be a cyclic code of even length $8$ over $R_1$. It is easy to check that $g(x)$ is self-reciprocal and $x^ip^*(x)=p(x)$, where $i=~deg~g(x)-deg~p(x)$.
		\item Cyclic codes of length $6$ over $R_1 = \F_2 + u \F_2 , u^2 = 1$: We have
		$$ x^6-1 = (x+1)^2(x^2+x+1)^2=f_1^2f_2^2 ~\text{over}~ \F_2.$$
		Let $C=\langle g(x), (1+u)a(x)\rangle, ~\text{where}~g(x)=f_1^2f_2, a(x)=f_2^2 $ be a cyclic code of length $6$ and type $4^12$ over $R_1$.
		We check that $g(x)$ and $a(x)$ are self-reciprocal. This code has
		16 codewords. These codewords are given in Table 4.
	\end{enumerate} 
\end{example}
\begin{center}
	{\bf Table 4.} A DNA code of length 6 obtained from the above code.\\~\\
	\begin{tabular}{ l  c   }
		\hline
		$AAAAAA$ &  $TTTTTT$ \\
		$ATATAT$ &  $TATATA$ \\
		$GGGGGG$ &  $CCCCCC$ \\
		$GCGCGC$ &  $CGCGCG$ \\   
		\hline
		
	\end{tabular}
\end{center}	

\section{Conclusion} 
In this paper, the algebraic structure of the ring $\F_2[u,v]/\langle u^2-1,v^3-v,uv-vu \rangle $ of odd length is studied. Cyclic codes is related to DNA codes and their relation is also studied. Reversible codes and reverse-complement codes related to cyclic codes are studied, respectively. Necessary and sufficient conditions for cyclic codes to have the DNA properties have been explored. Again we study binary image of cyclic codes over that ring via the Gray map.


\end{document}